\documentclass[fleqn,10pt]{wlscirep}
\pdfoutput=1
\usepackage[utf8]{inputenc}
\usepackage[T1]{fontenc}
\title{Memory-based optical polarization conversion in a double-$\Lambda$ atomic system with degenerate Zeeman states}

\author[1,2]{Yan-Cheng Wei}
\author[1,2]{Sheng-Xiang Lin}
\author[1,2]{Pin-Ju Tsai}
\author[1,3*]{Ying-Cheng Chen}
\affil[1]{Institute of Atomic and Molecular Sciences, Academia Sinica, Taipei 10617, Taiwan}
\affil[2]{Department of Physics, National Taiwan University, Taipei 10617, Taiwan}
\affil[3]{Center for Quantum Technology, Hsinchu 30013, Taiwan}
\affil[*]{chenyc@pub.iams.sinica.edu.tw}

\begin{abstract}
Optical memory based on electromagnetically induced transparency (EIT) in a double-$\Lambda$ atomic system provides a convenient way to convert the frequency, bandwidth or polarization of optical pulses by storing in one $\Lambda$ channel and retrieving in another. This memory-based optical converter can be used to bridge quantum nodes of different physical properties in a quantum network. However, in real atoms, each energy level usually contains degenerate Zeeman states and this may lead to additional energy loss, as have been discussed in our recent theoretical paper (Phys. Rev. A 100, 063843). Here, we present an experimental study on the efficiency variation in the EIT-memory-based optical polarization conversion in cold cesium atoms under the Zeeman-state optical pumping. The experimental results support the theoretical predictions. Our works provide quantitative knowledge and physical insight useful to the practical implementation of EIT-memory-based optical converter.      
\end{abstract}
\begin{document}
\flushbottom
\maketitle
%
%
\thispagestyle{empty}
\section*{Introduction}
Storage and retrieval of light pulses in atomic ensembles using the effect of electromagnetically induced transparency (EIT) in a three-level system have been intensively studied due to its important application as quantum memory for quantum information processing\cite{PhysRevLett.84.5094,PhysRevA.65.022314}. By controlling the intensity, frequency or direction of the control field during the retrieval process, the temporal width, frequency or propagating direction of the retrieved probe pulses can be manipulated\cite{PhysRevLett.88.103601, PhysRevA.69.035803, PhysRevA.72.053803,Chen:06}. With a four-level double-$\Lambda$ system, the wavelength of the retrieved probe pulses can be widely manipulated by turning on the control field of the second $\Lambda$ system during retrieval process\cite{RACZYNSKI2002149, PhysRevA.69.043801, Chen:06, Vewinger:07}. This can be served as a quantum frequency converter to interface different quantum systems in a quantum network. Furthermore, by turning on both control fields during retrieval, the stored atomic coherence can be simultaneously released into two separate photonic channels with the amplitude ratio controlled by the intensity ratio of the two control fields\cite{RACZYNSKI2002149,PhysRevA.72.053803,PhysRevA.69.043801,Park2016}. This can be served as a frequency-domain tunable beam splitter\cite{LI2005269} or two-color quantum memory\cite{PhysRevA.86.053827}.   

However, each energy level usually contains degenerate Zeeman states in real atoms and this may induce some complications in the memory-based optical conversion with a double-$\Lambda$ system. In a recent theoretical work\cite{tsai}, we have performed such a study and identified the two factors affecting the efficiency of the converted pulses. The first factor is the finite bandwidth effect of the optical pulses and the difference in the optical depth of the storage and retrieval channels. The second factor is the incompatibility between the stored ground-state coherence and the ratio of the probe and control Clebsch-Gordan coefficients, which lead to a nonadiabatic energy loss in the retrieved pulses\cite{RACZYNSKI2002149, Chen:06, PhysRevA.75.013812}. We obtained an approximate relation for the conversion efficiency.
Besides, we also numerically study the dependence of those two factors on the Zeeman population distribution, prepared by optical pumping\cite{tsai}.  

In this paper, we conduct an experimental study on the EIT-memory-based optical polarization conversion in a double-$\Lambda$ system with degenerate Zeeman states using cold cesium atoms. Specifically, we study the dependence of the conversion efficiency on various configurations, including the Zeeman population distribution prepared by optical pumping, the pulsed or continuous-wave probe, and the difference in optical dpeth for the storage and retrieval channel. The experimental observations support the experimental predictions. Our work provides useful quantitative knowledge and insight in the optical conversion based on EIT-memory. 

\section*{Results}
\subsection*{Theoretical Model}
\begin{figure}[t]
\centering
\includegraphics[width=0.8\textwidth,viewport=70 70 760 510,clip]{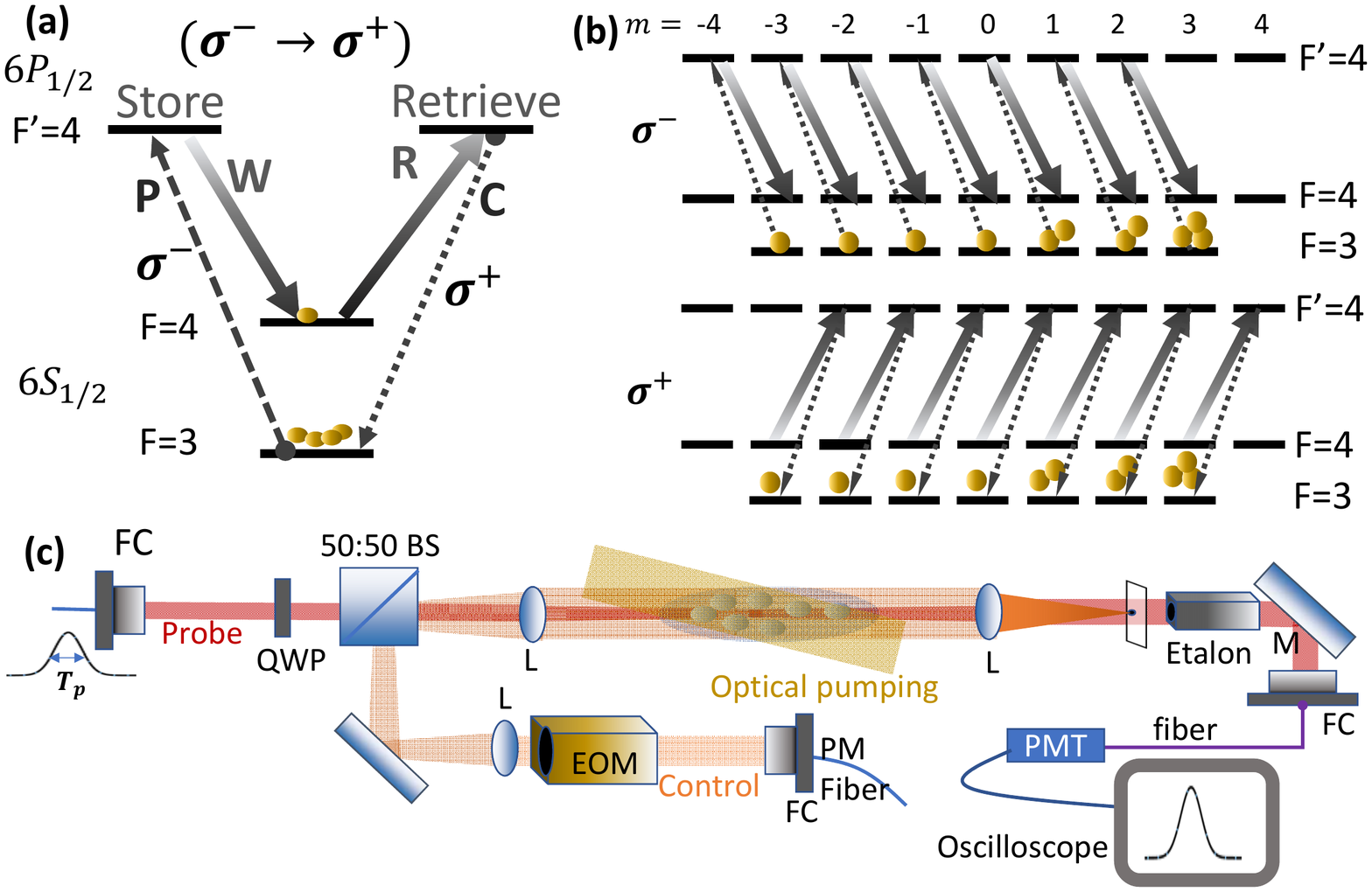}
\caption{(a) Relevant energy levels for $^{133}$Cs atoms and laser excitations. In this figure, we plot the case to store light into the spin-wave with a left-hand circularly polarized beam ($\sigma^-$ transition) and to retrieve with a right-hand circularly polarized beam ($\sigma^+$ transition). W,R,P,C denote writing control, reading control, probe, conversion fields, respectively. This case and the opposite case, i.e. storing by the $\sigma^+$ transitions and retrieving by the $\sigma^-$ transitions, are both implemented in the experiment. (b) The complete energy levels and laser excitations considering the degenerate Zeeman states. (c) Schematic experimental setup. BS: beam splitter; M: mirror; L: lens; FC: fiber coupler; QWP: quarter waveplate; PMT: photo-multiplier tube; EOM: electro-optic modulator; PM fiber: polarization-maintaining fiber;}
\label{Setup}
\end{figure}

Before mentioning the experiment, we briefly give a summary on the theoretical background\cite{tsai}. We consider the EIT-memory-based optical polarization conversion in a double-$\Lambda$ system with degenerate Zeeman states as shown in Fig.\ref{Setup}a and b for cesium $D_1$ line. In the $\sigma^- \rightarrow \sigma^+$ ($\sigma^+ \rightarrow \sigma^-$) conversion system, both the probe and writing control beams drive the $\sigma^-$ ($\sigma^+$) transitions in the storage channel and the conversion and reading control fields drive the $\sigma^+$ ($\sigma^-$) transitions in the retrieval channel. We focus our discussion on the relative conversion efficiency, $\xi^R_c$, which is defined as the energy ratio of the conversion pulse to that retrieved in the original storage channel. With this definition of $\xi^R_c$, the efficiency during the storage process has been normalized away. Based on Maxwell-Bloch equations and under certain approximations, we obtain \cite{tsai}
\begin{equation}
\begin{aligned}
\xi^R_c & =\xi_1(\eta)  \ \xi_2, \\
\xi_1(\eta) = \frac{\zeta_p(\eta)}{\zeta_c(\eta)}, & \ \
\xi_2 = \frac{\left|\sum_j p_j R_j^p R_j^c\right|^2}{\sum_j p_j {R_j^p}^2\sum_j p_j {R_j^c}^2},
\end{aligned}
\label{xi_I_zeeman}
\end{equation}
where $\zeta_p, \zeta_c$ are given by
\begin{equation}
    \begin{aligned}
    \zeta_p(\eta) =& \left[1+\frac{16ln2(\eta-\kappa)\eta}{\beta^2_w(L_w)}\frac{\sum_j p_j {R_j^p}^4/(a_{p,j}^2 \alpha_p)}{{(\sum_j p_j  {R_j^p}^2)}^2}\right]^{\frac{1}{2}}, \\
    \zeta_c(\eta) =& \left[1+\frac{16ln2(\eta-\kappa)\eta}{\beta^2_w(L_w)}\frac{\sum_j p_j {R_j^c}^4/(a_{c,j}^2\alpha_c)}{{(\sum_j p_j {R_j^c}^2)}^2}\right]^{\frac{1}{2}}
    \end{aligned}
    \label{defintions_Zeta}
\end{equation}
and
\begin{equation}
    \begin{aligned}
        \beta_w(L_w)=[1 + 16 ln2 \frac{\eta \kappa}{D_p}]^{\frac{1}{2}},
    \end{aligned}
    \label{defintions_Zeeman}
\end{equation}
where $R^p_j= \frac{a_{p,j}}{a_{w,j}}$, $R^c_j = \frac{a_{c,j}}{a_{r,j}}$, $\eta \equiv \frac{T_d}{T_p}$,   $\kappa \equiv \frac{T_w}{T_d}$, and $L_w=v_w T_w\simeq T_w L/T_d=L\kappa/\eta $. $v_w$ is the group velocity of the probe pulse in the storage channel. All subscripts $j$ denote the $j^{th}$ Zeeman sub-level, and $p_j$ denotes population in the $j^{th}$ ground state of the probe transition; $a_{(p,c,r,w)}$ denote Clebsch-Gordon coefficients for the probe, conversion, reading control, and writing control fields, respectively. $\alpha_p$ and $\alpha_c$ are the normalized optical depth of the probe and conversion transition without multiplied by the Clebsch-Gordan coefficients. The actual optical depths are $D_{(p;c)} = \sum_j{\alpha_{({p};{c})} a_{({p,j};{c,j})}^2}$. $T_d$ denotes group delay time of the slow light and can be expressed as $D_p \Gamma/\Omega_w^2$, where $\Omega_w$ is the Rabi frequency of the writing control field, and $\Gamma$ denotes the spontaneous decay rate; $T_w$ denotes the time when the writing control beam is switched off; $T_p$ denotes intensity full width at half maximum (FWHM) of the input probe pulse with a Gaussian waveform. 

\begin{figure}[t]
\centering
\includegraphics[width=0.9\textwidth,viewport=105 15 875 335,clip]{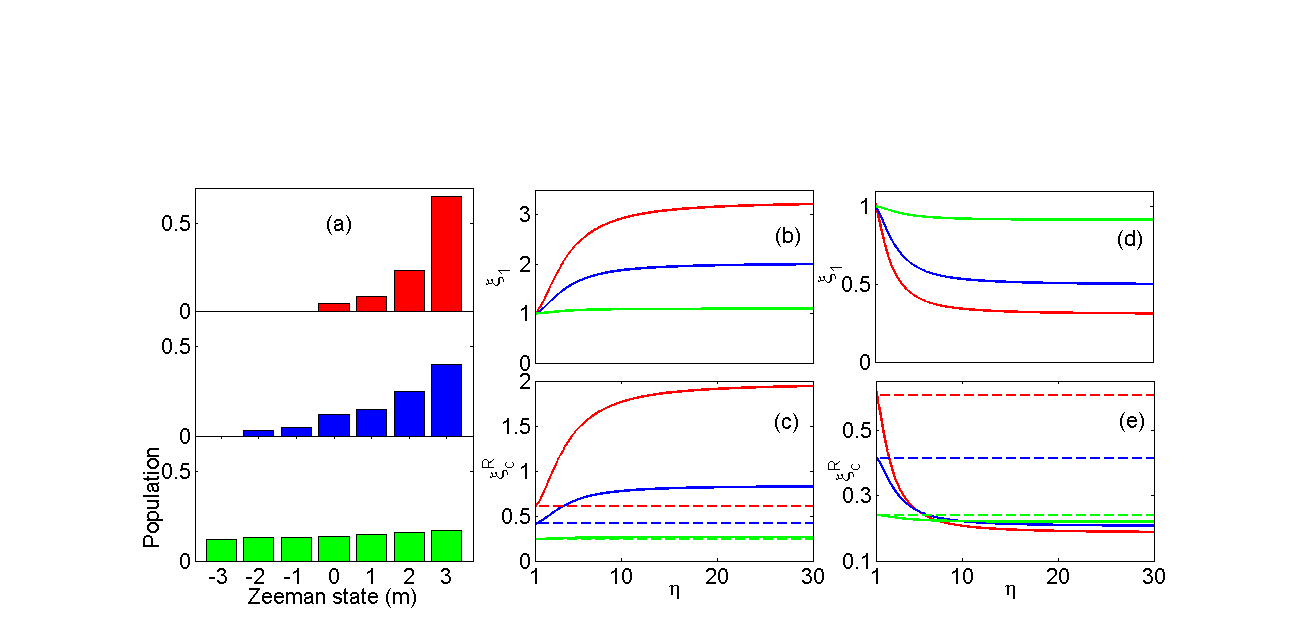}
\caption{Theoretical calculations based on Eq.\ref{xi_I_zeeman}. (a) Three assumed Zeeman population distributions. The Zeeman population changes from near isotropic distribution towards mostly concentrated in the rightmost Zeeman state from bottom to top. The optical depth for the probe driving the $\sigma^+$ ($\sigma^-$) transition increases (decreases) from the bottom population to the top one. The solid curves in (b, c) ((d, e)) are $\xi_1$ and $\xi_c^R$ versus $\eta$ for the $\sigma^- \rightarrow \sigma^+$ ($\sigma^+ \rightarrow \sigma^-$) conversion system. The color of the curves indicates the corresponding Zeeman population of (a) used in the calculation. The dash lines in (c) and (e) of the same color are the corresponding $\xi_2$. The parameters $\alpha_p=\alpha_c$=500 and $\kappa=1.1$.}
\label{Theoryplot}
\end{figure}

In Eq.\ref{xi_I_zeeman}, the relative conversion efficiency is determined by the product of two factors, which we term as the finite bandwidth factor ($\xi_1$) and the ground-state coherence mismatch factor ($\xi_2$)\cite{tsai}. $\xi_2$, ranging from zero to unity, stands for a nonadiabatic energy loss during the retrieval process when the population is distributed among several Zeeman states and a mismatch exists on the ratio $R_j^p/R_j^c$ for any of the $j^{th}$ sub-system  \cite{PhysRevA.75.013812,tsai}. The other factor $\xi_1$ is due to the finite bandwidth effect of the optical pulse and the difference in optical depth between the storage and retrieval channels\cite{tsai}. In the pulsed probe case, $\xi_1$ depends on the parameter $\eta=T_d/T_p=D_p\Gamma/(\Omega_w^2T_p)$. Because the EIT bandwidth of the storage channel $\Delta\omega_{EIT}\propto \Omega_w^2/(\sqrt{D_p}\Gamma)$ and the spectral bandwidth of the probe pulse $\Delta\omega_p \propto 1/T_p$, $\eta \propto \sqrt{D_p}\Delta\omega_p/\Delta\omega_{EIT}$, which is related to the ratio of the pulse bandwidth to EIT bandwidth\cite{PhysRevLett.120.183602}. This is why we call $\xi_1$ the finite-bandwidth factor. To have a clear picture of the trend based on Eq.\ref{xi_I_zeeman}, Fig.\ref{Theoryplot} (b,c) and (d,e) depict a theoretical plot for $\xi_1$ and $\xi_c^R$ versus $\eta$ for the $\sigma^- \rightarrow \sigma^+$ and $\sigma^+ \rightarrow \sigma^-$ conversion system, respectively. The three solid curves (red, blue and green) in Fig.\ref{Theoryplot}(b,c) and (d,e) correspond to the three Zeeman population distributions shown in Fig.\ref{Theoryplot} a of the same color. The three dash lines in (c,e) are the corresponding $\xi_2$, which are independent of $\eta$. Some trends are noted. First, $\xi_1$ is larger (smaller) than unity for the $\sigma^- \rightarrow \sigma^+$ ($\sigma^+ \rightarrow \sigma^-$) conversion system.
As discussed in Ref.\cite{tsai}, the retrieval efficiency is related to the delay-bandwidth product of the retrieved transition, which is dependent on the optical depth only. The retrieval channel with a larger optical depth has a higher retrieval efficiency. Considering the Clebsch-Gordan coefficients for cesium $D_1$ transition and the assumed Zeeman population, the $\sigma^- \rightarrow \sigma^+$ ($\sigma^+ \rightarrow \sigma^-$) conversion system corresponds to the conversion from a channel with a smaller (larger) optical depth to that with a larger (smaller) one. Second, the deviation of $\xi_1$ from unity is larger for the Zeeman population with a more concentrated distribution towards the $|m=3\rangle$ state. This is due to a larger contrast in the Clebsch-Gordan coefficient for the $\sigma^-$ and $\sigma^+$ probe transitions for the Zeeman state with a larger $m$ quantum number. In the case with isotropic Zeeman population, the overall transition dipole moment for the $\sigma^-$ and $\sigma^+$ probe transition are equal and thus $\xi_1$ is equal to unity. Third, $\xi_2$ increases with a more concentrated Zeeman distribution and is actually equal to unity when all population are in a single Zeeman state\cite{tsai}. Fourth, although $\xi_1 >1$ for the $\sigma^- \rightarrow \sigma^+$ system but $\xi_c^R(=\xi_1\xi_2)$ is larger than unity only if the Zeeman population is concentrated enough and if $\eta$ is larger than a certain value, as shown in Fig.\ref{Theoryplot}c. Even though the relative conversion efficiency is larger than unity, this situation occurs at a relatively large $\eta$ such that the absolutely conversion efficiency is significantly less than unity. Fifth, $\xi_1$ approaches unity when $\eta$ decreases for both $\sigma^- \rightarrow \sigma^+$ and $\sigma^+ \rightarrow \sigma^-$ conversion system. A smaller $\eta$ corresponds to a stronger writing control field and thus a wider EIT transparent bandwidth, compared to the spectral bandwidth of the probe pulse. The finite-bandwidth effect is thus less significant.

\subsection*{Experimental Details}
Our experiment is based on a cesium magneto-optical trap (MOT) with optically dense cold atomic media. Details of the MOT and experimental setup can be referred to Refs.\cite{PhysRevA.90.055401,PhysRevLett.120.183602}. The relevant energy levels and laser excitations are shown in Fig.\ref{Setup}a. For the storage channel, the writing control field and the probe field drive the $|F=4\rangle \rightarrow |F'=4\rangle$ and $|F=3\rangle \rightarrow |F'=4\rangle$ $\sigma^{-}$ transition of cesium $D_1$ line, respectively. For the retrieval channel, the reading control field and the conversion field drive the same transitions but with $\sigma^{+}$ polarizations, as shown in Fig.\ref{Setup}a. We compare this conversion system with that of the opposite driving case, i.e., storing with $\sigma^{+}$ transitions and retrieving with $\sigma^{-}$ transitions. Each hyperfine level in Fig.\ref{Setup}a contains multiple degenerate Zeeman states. The complete laser excitations are shown in Fig.\ref{Setup}b. In this experiment, we manipulate the polarization of the retrieved light but not its frequency. Although it is also possible to manipulate the frequency of the retrieved pulses by choosing another control transition during the retrieval process, our scheme allows us to concentrate our study on the relative efficiency change of the retrieved pulses without bothering by the systematic effects due to the detection of retrieved pulses of different wavelength. 

\begin{figure}[t]
\centering
\includegraphics[width=0.8\textwidth]{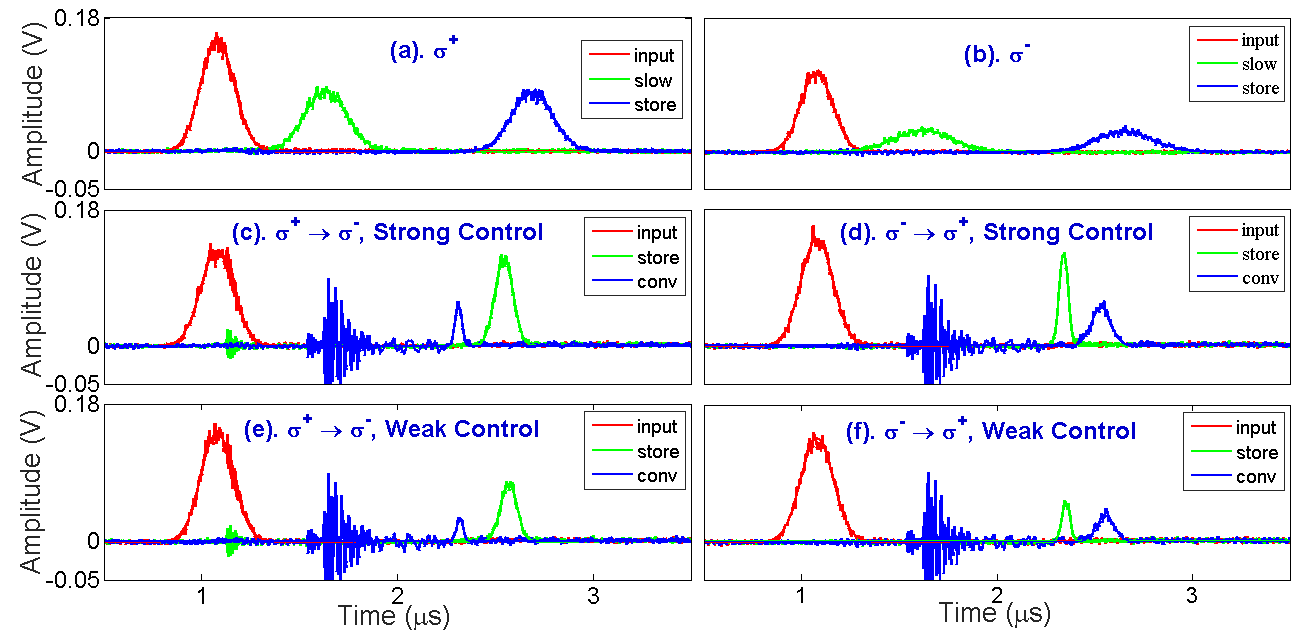}
\caption{Representative raw data. (a) and (b) depict the input (red), slowed (green), and stored and retrieved (blue) probe pulses in the same EIT channel: $\sigma^+$ and $\sigma^-$ channel, respectively. (c,e) and (d,f) depict the input (red), stored and retrieved in the original channel (green), and the converted (blue) pulses in the $\sigma^+ \rightarrow \sigma^-$ and $\sigma^- \rightarrow \sigma^+$ conversion system, respectively. The intensities of the writing control beams for (c) and (d) are stronger than those of (e) and (f), respectively. The Rabi frequencies for the writing control beam ($\Omega_w$) are (a) 4.03$\Gamma$, (b) 1.97$\Gamma$, (c) 3.11$\Gamma$, (d) 1.51$\Gamma$, (e) 2.11$\Gamma$, and (f) 0.93 $\Gamma$, respectively. The relative conversion efficiency ($\xi_c^R$) are (c) 17.4$\%$, (d) 93.8$\%$, (e) 15.4$\%$, and (f) 150.5$\%$, respectively. The high frequency noises appearing in (c,d,e,f) at roughly 1.7 $\mu$s come from the electronic noises due to the switching-on of the high voltage power supply for electro-optic modulator. The optical depths for the $\sigma^+$ and $\sigma^-$ EIT system are 389(41) and 52(3), respectively, where the quantity in the bracket is the 2$\sigma$ standard deviation.}
\label{Demo}
\end{figure}

The schematic setup of the experiment is shown in Fig.\ref{Setup}c. The probe beam doubly passes through one acoustic-optic modulator (AOM) for tuning its frequency and then passes through another AOM for shaping its temporal waveform into a Gaussian pulse. In the pulsed probe experiment, the full width at half maximum (FWHM) of the Gaussian probe pulse (denoted as $T_p$) is set to 200 ns. We also conduct the storage and retrieval experiment with a continuous-wave (CW) probe for a comparison with the pulsed probe case. More details of the CW probe experiment are described in Methods section. The probe beam is sent into the MOT cell via a polarization-maintaining fiber. The control beam passes through an electro-optic modulator (EOM) to quickly change its polarization within $\sim$10 ns after the storage and before the retrieval process. It then couples with the probe beam together through a beam splitter before entering the atomic clouds. The probe beam is focused to an intensity $e^{-2}$ diameter of $\sim$100$\mu$m around the atomic clouds while the control beam is collimated with a diameter of $\sim$1 mm. After coming out of the MOT cell, both beams pass through another lens and the control beam is focused while the probe beam is collimated. The control beam is blocked by a window with a black dot in the focal plane. The probe beam passes through three irises and an etalon filter, before coupled into a fiber and detected by a photomultiplier tube (Hamamatsu R636-10).

To control the ground-state coherence mismatch factor ($\xi_2$), which is sensitive to the Zeeman population distribution, we apply an optical pumping beam which drives the $D_2$ line $|F=3\rangle \rightarrow |F'=2\rangle$ $\sigma^+$ transitions\cite{PhysRevLett.120.183602}. It pumps atomic population toward the Zeeman states with larger magnetic quantum number $m$. We can control the Zeeman population distribution by the duration and/or intensity of the optical pumping beam. The Zeeman population can be determined by the microwave spectroscopy, described in the Methods section.

\subsection*{Experimental Observations}
We demonstrate the representative raw data in Fig.\ref{Demo}. In all these measurements, the optical pumping beam is applied for $20 \mu s$ to pump atoms towards the $|F=3, m=3\rangle$ state. Fig.\ref{Demo}a and Fig.\ref{Demo}b depict the input, slowed, and stored and retrieved optical pulses in the same $\Lambda$ system ($\sigma^+$ and $\sigma^-$, respectively). In order to achieve an optimized retrieval efficiency, we adjust the power of the control beam such that $\eta=T_d/T_p \approx$ 2.7\cite{PhysRevLett.120.183602}. Since the optical depth of the atomic clouds corresponding to the $\sigma^+$ channel is larger than that of the $\sigma^-$ channel under the optical pumping condition, the measured efficiencies of the slow light (67 \%) and stored light (66 \%) in the $\sigma^+$ EIT channel are larger than those (49 \%, 45 \%) of the $\sigma^-$ EIT channel\cite{PhysRevLett.120.183602}.

\begin{figure}[t]
\centering
\includegraphics[width=0.8\textwidth,,viewport=110 20 870 430,clip]{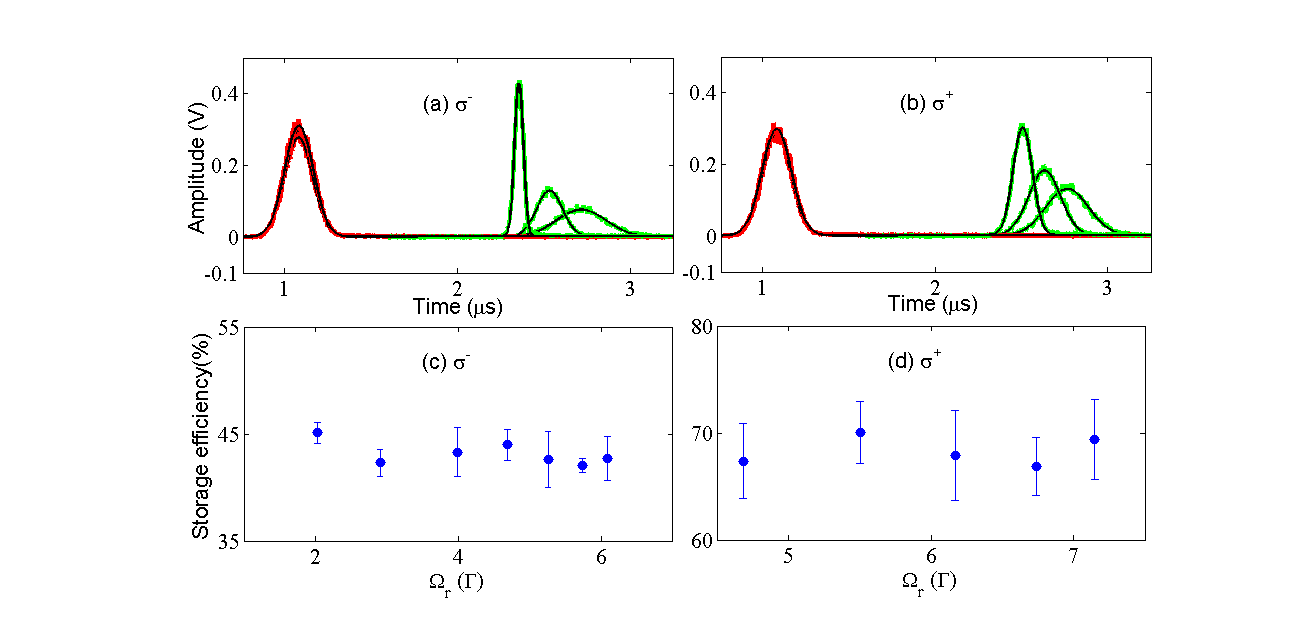}
\caption{(a) and (b) depict representative data of the stored and retrieved pulses (green) for some intensities of the reading control beam in the $\sigma^+$ and $\sigma^-$ channel, respectively. The red trace is the input pulse. All black solid lines are the Gaussian fits to the data. In (a), the Rabi frequency of the writing control beam $\Omega_r$ are 2.02, 2.90 and 6.09 $\Gamma$. In (b), the $\Omega_r$ are 4.68, 5.50, and 6.74 $\Gamma$. (c) and (d) depict the storage efficiency versus $\Omega_r$ for the $\sigma^-$ and $\sigma^+$ channel, respectively. The optical depth of the $\sigma^+$ and $\sigma^-$ EIT system are 349(32) and 47(3), respectively.}
\label{Ret}
\end{figure}

In the following, we present the data with polarization conversion. Fig.\ref{Demo}c and Fig.\ref{Demo}e (Fig.\ref{Demo}d and Fig.\ref{Demo}f) correspond to the case of storing via $\sigma^+$ channel ($\sigma^-$ channel) and retrieving via $\sigma^-$ channel ($\sigma^+$ channel). Both Fig.\ref{Demo}c and Fig.\ref{Demo}d store via stronger writing control field, compared to those of Fig.\ref{Demo}e and f which use weaker writing control field. The finite bandwidth factor $\xi_1$ should be less than unity in the $\sigma^+ \rightarrow \sigma^-$ conversion case, since it converts from the channel with a larger optical depth into that with a smaller one. Also, the coherence mismatch factor $\xi_2$ for this conversion case is smaller than unity because the Zeeman population is not completely in a single Zeeman state and the Clebsch-Gordan coefficients mismatch exist between the two $\Lambda$ systems. Thus, the relative conversion efficiency $\xi_c^R$ is smaller than unity based on Eq.\ref{xi_I_zeeman}, which agrees with both Fig.\ref{Demo}c and Fig.\ref{Demo}e. In the case of Fig.\ref{Demo}e with a weaker writing control field and thus a larger $\eta$, $\xi_1$ decreases further because of a stronger finite bandwidth effect and this leads to an even smaller $\xi_c^R$. In the opposite case of $\sigma^- \rightarrow \sigma^+$ conversion, the $\xi_1$ factor is larger than unity since it converts from the system of a smaller optical depth into that of a larger one. However, the $\xi_2$ factor is still smaller than unity and these two factors compete against each other. In the case with a stronger writing control regime (and thus a smaller $\eta$), the finite bandwidth effect is relatively less significant and $\xi^R_c$ is mainly affected by the coherence mismatch factor $\xi_2$. Thus, the overall $\xi^R_c$ is still smaller than unity, corresponding to the case of Fig.\ref{Demo}d. In the case with a weaker writing control beam (and thus a larger $\eta$), the finite bandwidth effect dominates and $\xi_1$ could be larger than unity. Therefore, the overall $\xi^R_c$ could be larger than unity for a large enough $\eta$, such as that of Fig.\ref{Demo}f. In this case, the efficiency of the converted light is even larger than that retrieved in the original channel. Although the relative conversion efficiency $\xi_c^R$ could be larger than unity, the absolute conversion efficiency is significantly lower than unity since $\xi_c^R >1$ occurs at a relatively large $\eta$ and in that case the efficiency of the storage process is relatively low.  

It should be noted that the pulse height and temporal width of the retrieval signal can be manipulated by the intensity of the reading control beam (with its Rabi frequency denoted as $\Omega_r$)\cite{PhysRevA.69.035803,PhysRevA.72.053803}. In an ideal three-level $\Lambda$-type EIT system, the retrieval efficiency should be independent of $\Omega_r$\cite{PhysRevLett.98.123601}. However, in real atom, there may exist nearby transitions such that the off-resonant excitation of the control beam may induce a $\Omega_r$-dependent multi-photon decay of the spin-wave\cite{PhysRevLett.81.3611,PhysRevLett.120.183602}.
Fortunately, the nearest off-resonant excitation of the control field for cesium $D_1$ line is quite far-detuned ($\sim$1.167 GHz for the $|6S_{1/2}, F=4\rangle \rightarrow |6P_{1/2}, F=3\rangle$ transition) such that the decay of the spin-wave due to the above-mentioned mechanism is negligible, at least for low $\Omega_r$\cite{PhysRevLett.120.183602}. To check if this is true, we measure the retrieval efficiency with different $\Omega_r$ for both the $\sigma^-$ and $\sigma^+$ EIT systems, as shown in Fig.\ref{Ret}. Within the experimental uncertainty, the retrieval efficiencies are roughly a constant for the range of used $\Omega_r$. This trend may not be valid if one implement the EIT-memory-based using the $D_2$ transition\cite{PhysRevLett.120.183602}. Because of this fact and in order to make the retrieval signal more clear, we choose a relatively strong reading control field to retrieve the signal, such as those shown in Fig.\ref{Demo}.  

\begin{figure}[t]
\centering
\includegraphics[width=\textwidth,viewport=20 200 600 400,clip]{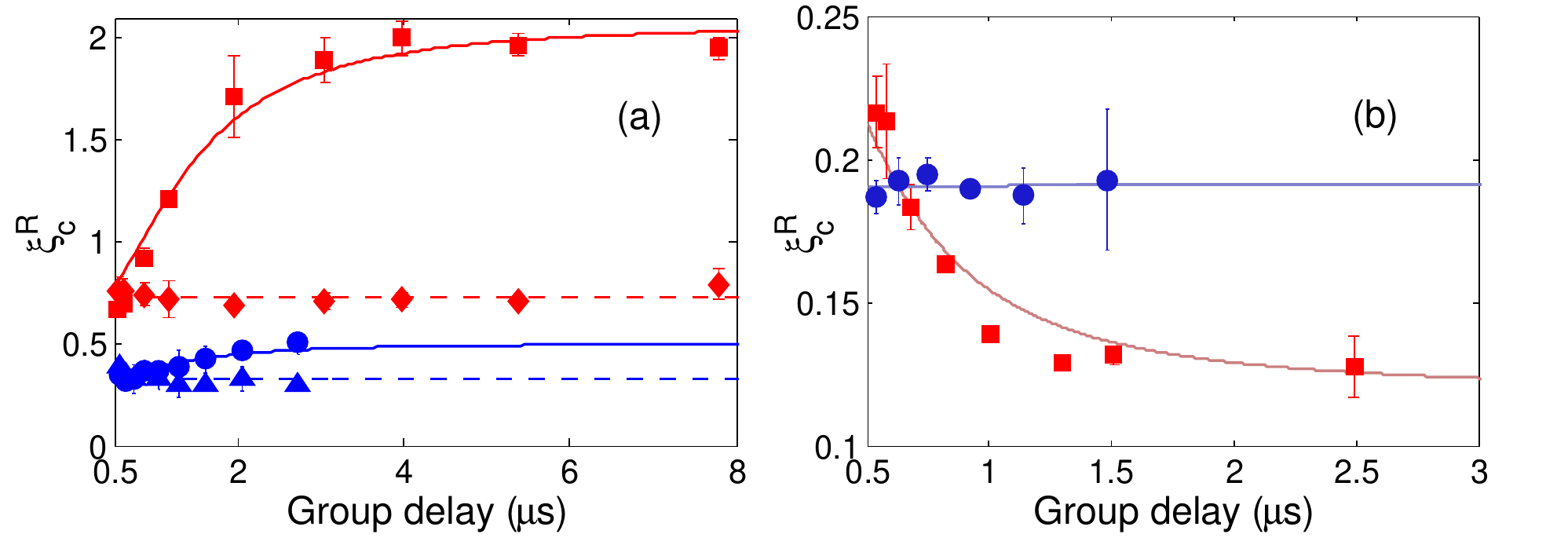}
\caption{Relative conversion efficiency $\xi^R_c$ versus group delay time of the storage process for (a) $\sigma^- \rightarrow \sigma^+$ and (b) $\sigma^+ \rightarrow \sigma^-$ conversion system. In (a), the red square (diamond) data corresponds to the pulsed (continuous) probe case with Zeeman optical pumping which pumps population towards the m=3 state. The blue circle (triangle) data corresponds to the pulsed (continuous) probe case without Zeeman optical pumping. The two dash lines correspond to the average values of $\xi^R_c$ for the continuous case. The two solid curves are the calculation by Eq.\ref{xi_I_zeeman} with $\alpha_p$=$\alpha_c$=1600, $\kappa$=1.5, and the population from $|F=3, m=3\rangle$ to $|F=3,m=-3\rangle$ states to be (0.56, 0.27, 0.13, 0.04, 0, 0, 0) and (0.25, 0.25, 0.15, 0.15, 0.10, 0.05, 0.05) for the case with and without Zeeman optical pumping, respectively. In (b), the red square (blue circle) data correspond to the pulsed probe case with and without optical pumping. The population from m=3 to -3 states are (0.52 0.24, 0.08, 0.04, 0.03, 0.02, 0.07) and (0.19, 0.17, 0.11, 0.10, 0.07, 0.17, 0.20), respectively. The two solid lines are the calculation by Eq.\ref{xi_I_zeeman} with $\kappa$=1.1 and $\alpha_p$=$\alpha_c$=400.
}
\label{all_data}
\end{figure}

To gain a more comprehensive picture, we conduct systematic measurements of $\xi^R_c$ versus $\eta$ by varying the $\Omega_w$ for various configurations. In the measurements, we all keep $T_p$= 200 ns. We consider both the $\sigma^- \rightarrow \sigma^+$ and $\sigma^+ \rightarrow \sigma^-$ conversion system, which corresponding to storing in a system with smaller optical depths and retrieving in another system with larger optical depths and the opposite. In each system, we also consider to vary the Zeeman population distribution by applying optical pumping beam or not. To explore the finite bandwidth effect, we also consider the storage and retrieval with a continuous-wave (CW) probe to serve as a reference for the pulsed probe case. 

Fig.\ref{all_data}a depicts the results for the $\sigma^- \rightarrow \sigma^+$ conversion system. The red square and blue circle are the data of the pulsed probe case with Zeeman optical pumping on and off, respectively. The red diamond and blue triangle are the data corresponding to the case of CW probe. It is evident that $\xi_c^R$ for the CW case almost show no dependence on the group delay time $T_d$. This is not surprising since the finite-bandwidth effect in the CW case is negligible and $\xi^R_c$ is only determined by $\xi_2$. Without optical pumping, Zeeman population is nearly isotropic and the $\xi_c^R$ shows a very weak dependence on $\eta$. With optical pumping, the Zeeman population concentrated towards Zeeman states with a larger $m$ and $\xi_c^R$ increases with $\eta$. At a large enough $T_d$, $\xi_c^R$ is larger than unity and approaches to a maximum value of $\sim$2 at large $T_d$. For the CW probe case, $\xi_c^R$ is larger for the case with optical pumping because the Zeeman population is more concentrated which favors a larger $\xi_2$. All these behaviors are similar to the theoretical predictions of Fig.\ref{Theoryplot} and have been explained in the theory section.               

We next conduct the corresponding experiments of the $\sigma^+ \rightarrow \sigma^-$ conversion system, with the results shown in Fig.\ref{all_data}b. Here, opposite to the previous case, the light is converted from the system with a large optical depth to that with a small one. The finite bandwidth factor $\xi_1$ is thus less then unity, which results in the decrease of $\xi^R_c$ when $T_d$ increases. This behavior is more significant when the optical pumping is on and the Zeeman population concentrated towards the Zeeman states with a larger $m$. $\xi^R_c$ becomes insensitive to $T_d$ when the optical pumping is off and the Zeeman population is more close to an isotropic distribution. The behaviors are similar to the theoretical predictions. 
 
\section*{Conclusion}
We conduct an experiment on EIT-memory-based optical polarization conversion in a double-$\Lambda$ system with cold cesium atoms. The dependence of the conversion efficiency on the degenerate Zeeman states and finite bandwidth effect have been studied. The experimental observations support the theoretical predictions. Our studies provide essential knowledge for the practical implementation of EIT-memory-based optical converter.     

\section*{Methods}

\subsection*{Measurement of the continuous-wave (CW) conversion efficiency}

\begin{figure}[t]
\includegraphics[width=1\textwidth,viewport=90 15 870 230,clip]{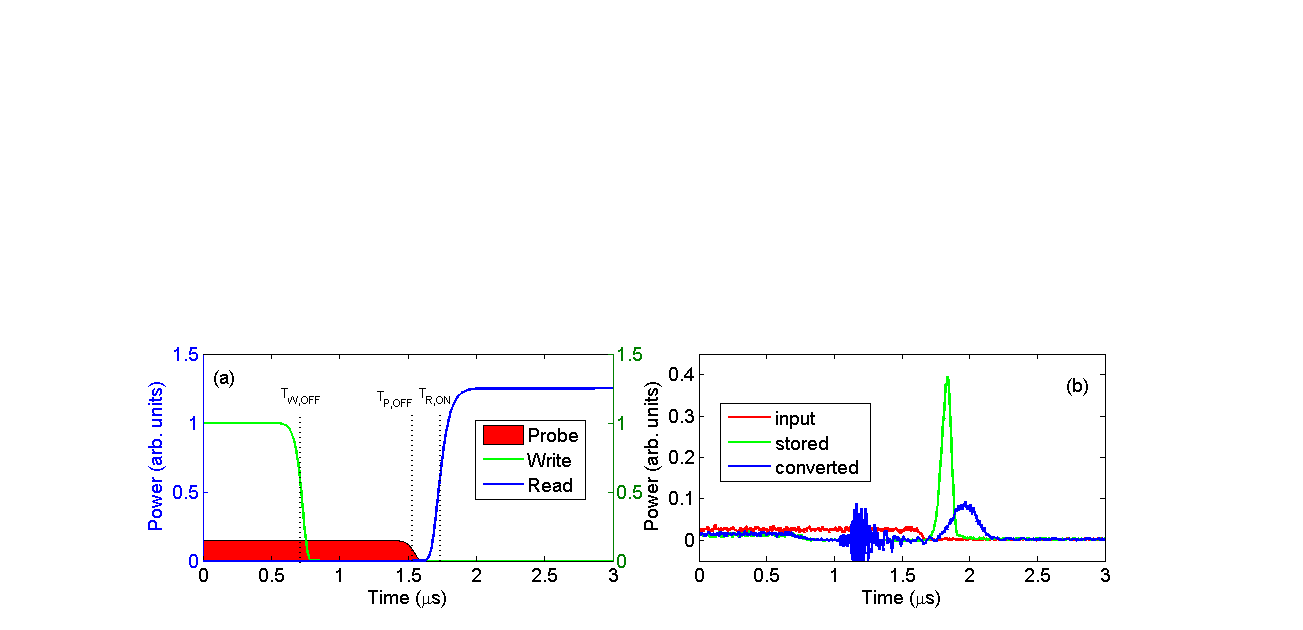}
\caption{(a) The schematic timing diagram for the probe (red), writing control (green) and reading control (blue) fields to implement the continuous probe storage and conversion measurement. The reading control beam could be that of the original $\Lambda$ channel or the converted $\Lambda$ channel. (b) One representative data for the input continuous probe (red), the retrieved probe fields in the original (green) and in the converted (blue) $\Lambda$ channel. The noises around 1.2 $\mu$s are due to the switching on of the high-voltage driver for EOM.}
\label{timeseq probe coupling}
\end{figure}

\begin{figure}[t]
\centering
\includegraphics[width=0.9\textwidth,viewport=135 15 910 420,clip]{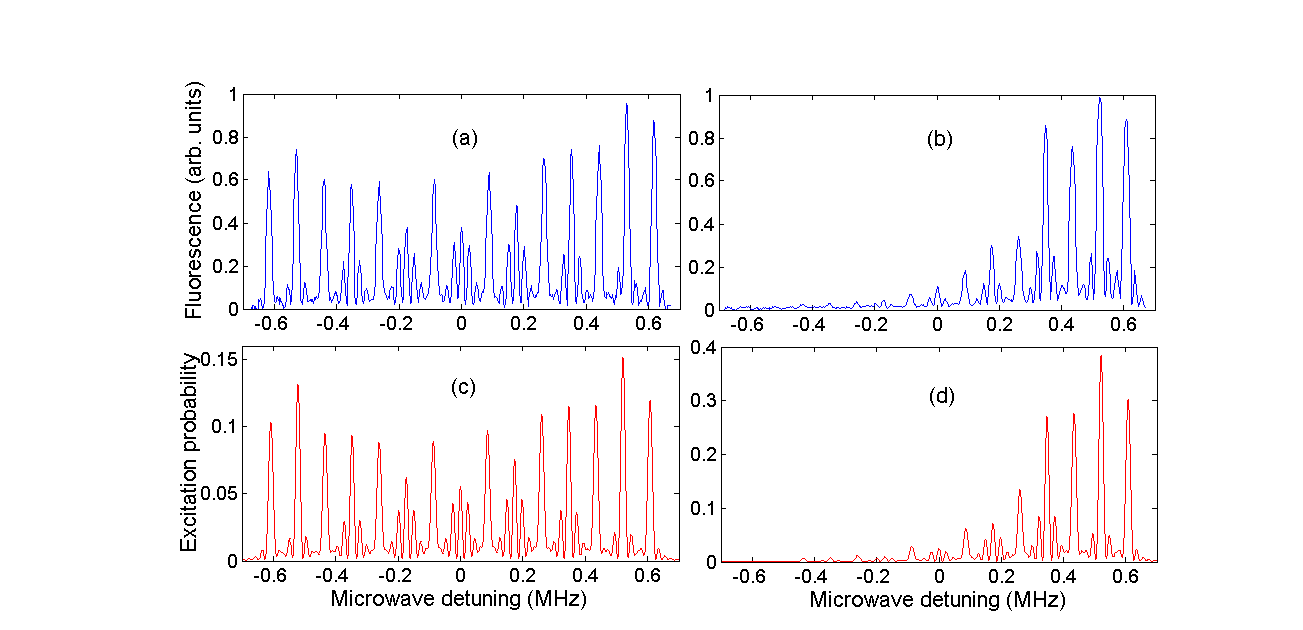}
\caption{(a) and (b) are the representative microwave spectrum with Zeeman optical pumping off and on, respectively. The x-axis is the microwave frequency minus 9192.63177 MHz, the hyperfine splitting between the $|F=3\rangle$ and $|F=4\rangle$ ground state. (c) and (d) are the calculated microwave spectrum corresponding to (a) and (b) with population from $|m=3\rangle$ to $|m=-3\rangle$ to be (0.15, 0.16, 0.16, 0.14, 0.13, 0.13, 0.13) and (0.38, 0.38, 0.15, 0.06, 0.02, 0.01, 0), respectively. In the calculation, the magnetic field is 248 mG. The microwave Rabi frequencies for the $\pi$ transitions are 3.936$\times 10^5$ Hz, multiplied by the Clebsch-Gordon coefficients. The Rabi frequency ratio for the microwave $\sigma$ to $\pi$ transition is 0.36, which is related to the angle between the microwave magnetic field to the applied DC magnetic field.}
\label{microwave}
\end{figure}

To implement the storage and conversion of the continuous-wave (CW) probe field, we firstly turn on both the writing control beam and the probe beam. The probe power is kept at a constant value. The writing control beam is switched off at a given time (denoted as $T_{W,OFF}$). After a certain time, the probe beam is also turned off (denoted as $T_{P,OFF}$). This ensure that the atomic ensemble is filled by a certain portion of the probe beam of constant intensity such that its behavior is more like the CW case. At time $T_{W,OFF}$, the portion of probe beam that enters the atomic medium will be converted to and stored as the spin-wave. Other parts of the probe beam that enter the medium later than $T_{W,OFF}$ will be absorbed. At a time (denoted as $T_{R,ON}$) later than $T_{P,OFF}$, the writing control beam, either of the original $\Lambda$ system or the converted one, is turned on to retrieve the probe pulse in the original $\Lambda$ system or in the converted $\Lambda$ system, respectively. The schematic timing diagram is shown in Fig.\ref{timeseq probe coupling}a. One representative data of such measurement is shown in Fig.\ref{timeseq probe coupling}b. The ratio of the retrieved probe energy in the converted $\Lambda$ system (blue) to that in the original $\Lambda$ system (green) is the relative conversion efficiency of the CW case. 

\subsection*{Determination of the Zeeman population by microwave spectroscopy}
To determine the Zeeman population distribution, we conduct the microwave spectroscopy. In the measurement, we apply a magnetic field of $\approx 248 mG$ and turn on a microwave pulse of 50 $\mu$s duration through a horn antenna to drive atoms from the $|6S_{1/2}, F=3\rangle$ Zeeman states into the $|6S_{1/2}, F=4\rangle$ Zeeman states. The trapping beams are then turned on with its frequency jumped to resonance of the $|6S_{1/2}, F=4\rangle \rightarrow |6P_{3/2}, F'=5\rangle$ cycling transition. One CCD camera is then used to collect the atomic fluorescence for 70 $\mu$s. The microwave frequency is scanned through 9.192 GHz and the timing sequence repeat to get the spectrum. There are 15 major lines appear in the spectrum due to the Zeeman shifts with an example shown in Fig.\ref{microwave}a and b with the Zeeman optical pumping off and on, respectively. Some lines are split into several sub-lines due to the relatively strong microwave field used for the experiment. Because the total trapping laser intensity is very strong ($\sim$ 100 $mW/cm^2$ corresponding to an on-resonance saturation parameter of $\sim$37 with respect to the isotropic saturation intensity of 2.706 $mW/cm^2$)\cite{Steck98cesiumd}, the fluorescence rate for each Zeeman sublevel is almost saturated to $\Gamma/2$ and the relative intensities of the spectral peaks are determined by the Zeeman population and the relative strength of the microwave transition only. By varying the microwave Rabi frequency, the Rabi frequency ratio between the microwave $\pi$ to $\sigma$ transition and the Zeeman population, the calculated microwave spectrum resembles the experimental spectrum very well, as shown in Fig.\ref{microwave}c and d. As shown in Fig.\ref{microwave}d, the $\sigma^+$ optical pumping beam do pump the Zeeman population towards the Zeeman states with larger magnetic quantum number $m$.

\bibliography{sample}



\section*{Acknowledgements}
This work was supported by the Ministry of science and technology of Taiwan under grant numbers MOST 108-2112-M-001-030-MY3 and MOST 108-2639-M-007-001-ASP. We also thank the supports from National Center for Theoretical Sciences of Taiwan under ECP1 program and Center for Quantum Technology of Taiwan. 

\section*{Author contributions statement}

YC W. and SX L. conducted the experiment and analysed the data. YC W. and PJ T. established the theory. YC W. and YC C. wrote the manuscript. All authors reviewed the manuscript. 

\section*{Additional information}

\textbf{Accession codes} (where applicable); 
\textbf{Competing interests} 
The authors declare no competing interests.

\end{document}